\documentclass[12pt]{article}
\usepackage{amsfonts,amsmath,amssymb,amscd}
\input epsf
\begin{document}

\vspace*{2.3cm}

%New title
\centerline{{\sc A CLASS OF DEDUCTIVE THEORIES THAT CANNOT BE DETERMINISTIC:}}
 \centerline{{\sc CLASSICAL AND QUANTUM PHYSICS ARE NOT DETERMINISTIC}}

\bigskip
\centerline{{\sc Iegor Reznikoff}}
\centerline{Professor Emeritus, D\'epartement de Philosophie,}
\centerline{Universit\'e de Paris-Ouest, 92001 Nanterre, France}

\smallskip
\centerline{E-mail: dominiqueleconte@yahoo.fr}

\bigskip

%%%%%%%%%%%%new text1%%%%%%%%%%%%%%%%%%%%%%%%%%%%%
The problem of the determinism of Quantum Mechanics has been a main one during the 
20th century. At the same time, in the context of Logic and Set Theory, the importance of 
ancient paradoxes as well as the appearance of many new ones, has shed light on and deeply 
influenced the foundations of Mathematics and somehow of Physics. But, strangely, 
concerning Physics, a paradox which we call the Memory Paradox has remained yet 
undiscovered, despite its simplicity and remarkable consequences, mostly in Physics and 
surprisingly in classical Physics that appear to be non deterministic, contrary to the general 
belief since Newton, Laplace, etc.. The non determinism of Quantum Physics follows without 
any supplementary hypothesis. This paper extends a previous one [6].

\bigskip

\centerline {{\sc 1. Introduction}}

\medskip

Among the main paradoxes that influenced the foundations of mathematics, let us mention Russel's Paradox (\textit{The Principles of Mathematics}, 1903) which showed the necessity of an 
axiomatic approach to set theory, and the most ancient, so called Liar Paradox
("This sentence is false"), which has led to the famous Godel's incompleteness theorems. Concerning 
Physics, there is, however, a paradox - we call it the Memory Paradox - that, curiously, has 
remained unknown, despite its simplicity and deep consequences. It is the fact that memory 
cannot completely predict its own future state.
The idea is simple. If the memory $m(t)$ at time~$t$ knows its future state~$m(t'$), at time $t' > t$, then 
$m(t')$ must be contained in some way in~$m(t)$.
But since the future memory always contains 
previous ones (it is the main property of memory), $m(t)$ is contained in $m(t')$ which is 
contained in $m(t)$; we have thus a circularity or vicious circle, typical of paradoxes. That this 
Memory Paradox leads to a contradiction is shown in the next sections.

\bigskip
%%%%%%%%%%%%old text%%%%%%%%%%%%%%%%%%%%%%%%%%%%%

\centerline {{\sc 2. Definitions}}

\medskip
By a {\it deductive theory}, we mean a theory that is based on (not necessarily completely formal) logical reasoning, as e.g. mathematics, logic itself of course, or more casually physical science, theoretical or applied. If $F$ is the theory and $\vdash$  the symbol of deduction, $F \vdash  A$  means that $A$ is deducible from (or can be proved in) the theory $F$.

By a {\it deterministic theory}, we mean a deductive theory in which a time variable $t$ is defined, $t$ belonging as usual to an interval of real numbers (however an ordered set is sufficient here), and any true event $E(t)$ expressible in the theory and occurring at time $t$ can be deduced 'in advance' in the theory, i.e., there exists a  $t' < t$  such that
\begin{equation}        
F(t')\vdash  E(t),
\end{equation}
where in $F$ appear only events that occurred at time $\leq t'$ (this can be formulated precisely in logical terms, but this is not necessary here). In such case the event $E(t)$ is said to be predicted or (pre)determined at time $t'$ by theory $F$.

By an {\it observing machine} $M$ with {\it memory} based on a deductive theory $F$ we mean a machine (e.g., a Turing machine) a computer or a human brain that can observe (or read) any finitely coded sentence or formula (expressible in $F$) that appears on a tape or in the coded observational field of the machine; moreover, the machine stores this sentence or a part of it in its memory, i.e., the finite set of already stored sentences since a given time (the machine, of course, can observe only a finite number of times). In particular, the machine $M$ can observe the second member of formula (1) above relative to the event $E(t)$ as soon as the formula has appeared among (or in the list of) deducible formulas in $F$, i.e., at time $t' < t$ , and stores the predicted event $E(t)$ in its memory.

Let us write $M(t, x(t), m(t) )$ for ``at time $t$ $M$ observes $x(t)$ and stores it in its memory 
$m(t)$". By the definition above, the memory obviously has the following property:
\begin{equation}    
\text{if} \quad t' < t \quad \text {then} \quad  m(t') \subset m(t).
\end{equation}
This corresponds to the very meaning of the notion of memory.

\bigskip

\centerline{{\sc 3. Main Results}}

\medskip
{\bf Theorem.}  {\it The existence of an observing machine with memory based on a deductive theory $F$ is inconsistent with the theory $F$ being deterministic. }

\medskip

{\bf Proof.}  Suppose the contrary. Since the theory $F$ is deterministic, in particular in what concerns $M$, at some time $t' < t$  the behavior of the machine $M$ at time $t$ is predetermined in the theory, i.e., 
\vspace{-1ex}
%%\null
%%.
%%\vskip 2.3cm

\noindent $F(t') \vdash  M(t, x(t), m(t) )$. Then already at time $t'$, the machine $M$ observes the predicted event and stores it in its memory $m(t')$, thus 
\begin{equation}
M(t, x(t), m(t) ) \in  m(t').
\end{equation}
Now, when later on, time $t$ indeed occurs, if the deterministic prediction is correct, the expected event $M(t, x(t), m(t) )$ should occur as well, but since $t' < t$ , because of (2) we have 
$m(t') \subset m(t)$, and therefore, $M(t, x(t), m(t) )\in  m(t)$, which is impossible since $m(t)$ 
is finitely coded.

\medskip

{\bf Commentary.}
What we need is that the interval of time between $t'$ and $t$ be large enough for $M$ to have observed and stored (possibly at time $t''$ with $t' < t'' \leq t$) the predicted sentence 
$M(t, x(t), m(t) )$ before (or at  the latest at) time $t$. Technically for such simple a task, a machine nowadays needs a very small fraction of time, while a deterministic theory, if all needed information is available, should calculate and predict an event, in principle, ``sufficiently'' in advance. Otherwise, if there is no time to know a predicted result before it happens, there is of course no prediction and the determinism would have no more value than the religious belief that ``everything is already written'' in some Book for ever unknown. In any case, if we accept determinism in the ordinary meaning of the word so that the predicted event can be recorded as such, the theorem shows that it is not compatible with observation and memory. 

\medskip

{\bf Corollary 1.}  {\it In the deterministic part of classical physics, an observing machine with memory cannot be defined $($and therefore constructed$)$.}

\medskip
Now, since an observing machine with memory certainly exists in our physical world, e.g., as the brain of human being or, more elementarily, as a modern computer, we have

\medskip
{\bf Corollary 2.}  {\it Physics as a whole cannot be completely deterministic.} 

\medskip
%%%%%%%%%%%%new text2%%%%%%%%%%%%%%%%%%%%%%%%%%%%%

Now, if we restrict the reasoning to classical physics,
the same conclusion appears if we 
assume the classical determinism. Indeed, because of the paradox,
a deterministic classical Machine cannot predict its own future memory,
refuting the famous idea of Pierre-Simon 
Laplace of a 'Demon' who, knowing all the data of the universe at time t,
could predict any future event (at time $t' > t$) [3].
The Demon could predict any event except its own future 
memory.
Even if we suppose that all functions are continuous, the prediction is not possible.
The  problem is that the initial data needed for the calculus (to be made by the
Machine to predict its future) are not well defined; indeed, these data should already integrate the fact that the 
Machine is going to perform the calculus on these data, so that the definition of the set of data 
contains a circularity. Of course, the Machine can calculate and predict events in which the 
Machine is not involved.
\medskip

{\bf Corollary~3.} \textit{Classical physics are not deterministic}.

\medskip
This result is different from the results starting from Hadamard [2] and Poincar\'e [4] which 
concern deterministic but not (practically) predictive dynamic situations related to what is 
now called chaos. Corollary~3 refutes the theoretical determinism of classical Physics while  
Chaos Theories do not.
The question appears whether this non determinism is meaningful only on the macroscopic 
level because of the concept of a machine, or whether it is true also at the quantum level.
\medskip

{\bf Corollary 4.} \textit{Quantum Mechanics, if consistent, are not deterministic}.

\medskip
   {\bf Proof.}  In [5], we have given a formal logical proof
(improving, by its purely logical 
approach, the Conway - Kochen Free Will Theorem [1]) that if an observer of spin 
components is free in this observation, then Quantum Mechanics are not deterministic; but 
our main theorem here proves that the behaviour of an observer or machine with memory cannot be predetermined, i.e., is necessarily free.

\medskip
It is important to remark that, in this proof, no supplementary conditions (as e.g. the freedom 
of the observer) are needed. What is needed is not 'freedom' but that the observer has memory 
and this, of course, is true. 

If observed, Nature cannot be deterministic; actually, when observed, Nature is observing 
itself, while certainly it includes memory, and this, as we have shown, yields the non-
determinism of both classical and quantum Physics. Extending Einstein's famous saying, we 
ought to say: God doesn't play dice, but lets the dice play.

\bigskip

\centerline {{\sc References}}

\vspace{-1ex}
\begin{itemize}
\item[1.] J.~Conway and S.~Kochen, \textit{The Strong Free Will Theorem},
AMS, vol 56/2,
pp. 226--232, Providence, RI, February 2009.

\item[2.] J.~Hadamard, ``Les surfaces \`a courbures oppos\'ees
et leurs lignes g\'eodesiques,''
Journal de Math\'ematiques pures et appliqu\'ees, t. IV, 1898, pp. 27--73;
{\OE}uvres, t. II, pp.729--775, CNRS Editions, Paris, 1968.

\item[3.] P.-S. Laplace, \textit{Essai philosophique sur les probabilit\'es}, pp. 4--5, Paris, 1814.

\item[4.] H.~Poincar\'e, \textit{Sur le probl\`eme des trois corps
et les equations de la dynamique},
Acta Mathematica, t. 13, pp. 1--270, 1890.

\item[5.] I.~Reznikoff, \textit{A Logical Proof of the Free Will Theorem}, arXiv: 1008.3661v1 [quant-ph], 21, Aug 2010 (http://arxiv.org/abs/1008.3661).

\item[6.] I.~Reznikoff, \textit{A Class of Deductive Theories that Cannot Be Deterministic},
arXiv: 1203.2945v1 [physics.gen-ph], 13 Mar 2012
(http://arxiv.org/abs/1203.2945).
\end{itemize}

\end{document}